\documentclass[fleqn,usenatbib]{mnras}

\usepackage[T1]{fontenc}

\DeclareRobustCommand{\VAN}[3]{#2}
\let\VANthebibliography\thebibliography
\def\thebibliography{\DeclareRobustCommand{\VAN}[3]{##3}\VANthebibliography}

\usepackage{graphicx}	
\usepackage{amsmath}	
\usepackage{amssymb}	
\usepackage{newtxtext,newtxmath}
\usepackage{graphicx}
\usepackage{txfonts}
\usepackage{color}
\usepackage{stfloats}
\usepackage{amsmath}
\usepackage{multirow}
\usepackage{booktabs,caption}
\usepackage[flushleft]{threeparttable}
\usepackage{array}
\usepackage{epsfig}
\usepackage{amsmath}
\usepackage{subfigure}
\usepackage{natbib}
\usepackage{tablefootnote}
\usepackage{longtable}
\usepackage{graphicx}
\usepackage{epstopdf}
\usepackage{booktabs}
\usepackage{txfonts}
\usepackage[utf8]{inputenc}%
\usepackage{graphicx}
\usepackage{txfonts}
\usepackage{stfloats}

\title[Rotational spectroscopy of ethynylnaphthalenes]{Laboratory rotational spectroscopy and astronomical search of ethynyl substituted naphthalene}

\author[Carlos Cabezas et al.]{
Carlos Cabezas,$^{1}$\thanks{E-mail: carlos.cabezas@csic.es (CC)}
Isabel Pe\~na$^{2}$\thanks{E-mail: ipencal@qf.uva.es (IP)}
and Jos\'e Cernicharo$^{1}$
\\
$^{1}$Departamento de Astrofísica Molecular, Instituto de Física Fundamental, Consejo Superior de Investigaciones Científicas, \\
C/ Serrano 121, 28006 Madrid, Spain \\
$^{2}$Departamento de Qu\'imica F\'isica y Qu\'imica Inorg\'anica, Facultad de Ciencias, Universidad de Valladolid, Paseo de Bel\'en 7,
Valladolid 47011, Spain
}

\date{Accepted 2022 December 13. Received 2022 December 11; in original form 2022 September 9}

\pubyear{2022}

\begin{document}
\label{firstpage}
\pagerange{\pageref{firstpage}--\pageref{lastpage}}
\maketitle

\begin{abstract}

The recent interstellar detection of cyanonaphthalenes bring interest in related aromatic molecular species that could be present in similar astronomical environments. In this context, ethynyl derivatives of naphthalene are promising candidates to be observed in the Taurus Molecular Cloud (TMC-1), where cyanonaphthalenes together with cyano- and ethynyl- derivatives of cyclopentadiene and benzene have been detected. To enable the interstellar searches for ethynyl derivatives of naphthalene, their pure rotational spectra need to be investigated in the laboratory. We have observed for the first time the rotational spectra of 1- and 2-ethynylnaphthalene species using a broadband Fourier-transform microwave spectrometer operating in the 2-8 GHz frequency region. Accurate spectroscopic parameters are derived from the analysis of the experimental spectra, allowing for reliable predictions for astronomical searches. Our searches in TMC-1 for both isomers provide upper limits for the abundances of these species.

\end{abstract}




\section{Introduction}

Aromatic molecules are dominant in the chemistry occurring on Earth, many playing important roles in biological systems, i.e., nucleobases and their derivatives, nucleosides. They are ubiquitous also in the Universe, where around 20\% of all interstellar carbon, is thought to be incorporated into polycyclic aromatic hydrocarbons (PAHs) \citep{Tielens2008}. PAHs are proposed as carriers of the infrared emission features that dominate the spectra of most galactic and extragalactic sources \citep{Leger1984,Allamandola1985,Tielens2008}. However, the formation processes of these species have been poorly understood due to the inability to detect individual molecules. The first detection of an aromatic molecule in space was in 2001, when benzene was identified in the protoplanetary carbon-rich star CRL618 \citep{Cernicharo2001} through its infrared spectral signature. However, despite their expected ubiquity, further astronomical identification of aromatic molecules had proven elusive for almost two decades, until benzonitrile was detected in 2018 by \citet{McGuire2018} in the dense molecular cloud TMC-1 through its rotational spectrum. Very recently, four new aromatic molecules have been discovered in TMC-1 by their rotational spectra, thanks to the high level of sensitivity available with the Green Bank Telescope and the Yebes 40m telescope and the observing projects GOTHAM\footnote{\textbf{G}BT \textbf{O}bservations of \textbf{T}MC-1: \textbf{H}unting \textbf{A}romatic \textbf{M}olecules} \citep{McGuire2018} and QUIJOTE\footnote{\textbf{Q}-band \textbf{U}ltrasensitive \textbf{I}nspection \textbf{J}ourney to the \textbf{O}bscure \textbf{T}MC-1 \textbf{E}nvironment} \citep{Cernicharo2021a}. These new molecular species are the two isomers of the cyano derivatives of naphthalene, 1- and 2-cyanonaphthalene \citep{McGuire2021} and the first pure PAH discovered so far, indene \citep{Cernicharo2021b,Burkhardt2021} together with its cyano derivative \citep{Sita2022}.

Indene, despite its large molecular size, is exceptionally abundant in TMC-1, just five times less abundant than the ubiquitous cyclic hydrocarbon $c$-C$_3$H$_2$ \citep{Cernicharo2021b}. The detection of cyano derivatives of indene serves to directly compare the ratio of a pure hydrocarbon PAH to its cyano substituted counterpart in observations versus those predicted by astrochemical models. \citet{Sita2022} posit that cyano derivatives of PAHs indeed can be used as an excellent observational proxy for their hydrocarbon counterparts for constraining models, at least within a factor of a few. Based on this, benzene and naphthalene are also expected to be very abundant in TMC-1, although they were supposed to be formed primarily in the circumstellar envelopes of evolved stars  \citep{Tielens2008}. In this scenario and due to the extraordinary level of sensitivity of the current TMC-1 line surveys, it is probable that further PAHs are awaiting detection. However, most of PAHs are very high symmetry molecules, which implies that they lack of permanent electric dipole moment preventing their radio and millimeter detection. As such, two alternatives can be followed to improve our understanding on the chemical processes leading to these molecular species. The first one is the identification of the intermediate species that are probably behind the in-situ bottom-up synthesis of PAHs in these environments \citep{Agundez2021}, and the second one, the direct detection of polar derivatives of the target aromatic molecules, i.e. benzonitrile and cyanonaphthalenes \citep{McGuire2018,McGuire2021}. New results in any of this research lines will certainly help in obtaining a complete chemical network to explain the molecular chemistry of aromatics and PAHs in TMC-1.

As mentioned above, the detections of 1- and 2-cyanonaphthalene species in TMC-1 \citep{McGuire2021} suggest that naphthalene should be quite abundant species in TMC-1. \citet{Cernicharo2021c} reported the abundances for -CN and -CCH derivatives of cyclopentadiene and benzene and their results indicate that the -CCH derivatives have larger abundances for both systems. \citet{McGuire2021} discussed the formation routes for the cyanonaphthalenes and one of the considered reactions is that between naphthalene molecule and CN radical. If we consider an analogue reaction to form the ethynyl derivatives of naphthalene, then these species could reach higher abundances than the cyano derivatives, as it occurs for cyclopentadiene and benzene, due to the large abundance of CCH radical compared to CN radical. It is then straightforward to think that ethynylnaphthalene can be considered as a good candidate to be detected through radio astronomy observations in TMC-1. In this work, we have employed Fourier transform microwave (FTMW) spectroscopy to obtain and assign the rotational spectra of the two isomers of ethynyl-substituted naphthalene, 1-ethynylnaphthalene (1-ETN) and 2-ethynylnaphthalene (2-ETN). The molecular constants for both species are accurately determined, in good agreement with previous theoretical predictions \citep{Ye2022}, and provide sufficient data for interstellar searches in TMC-1.

\begin{figure}
\centering
\includegraphics[angle=0,width=0.5\textwidth]{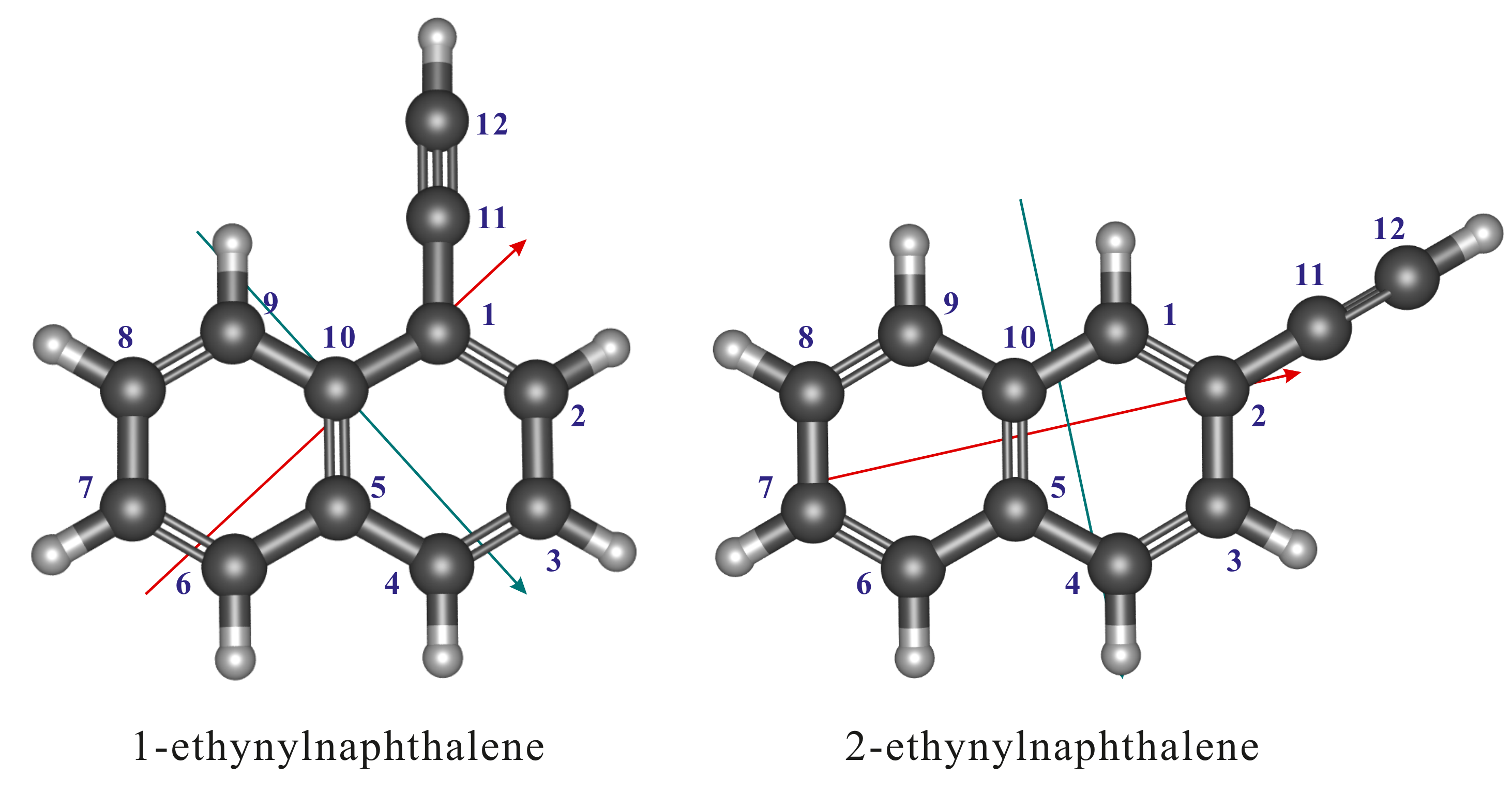}
\caption{Molecular structures and carbon atom labelling of the two isomers of ethynylnaphthalene. The $a$ (red) and $b$ (green) principal inertial axes are shown. The $c$ axis is perpendicular to the molecular plane.} \label{moles}
\end{figure}

\section{Experimental}

The rotational spectra of the 1-ETN and 2-ETN species were observed using a chirped-pulsed Fourier-transform microwave spectrometer operating in the frequency region of 2-8 GHz \citep{Pena2020,Cabezas2021}. Samples of 1-ETN (stated purity 97\%) and 2-ETN (stated purity >95\%) were purchased from Sigma Aldrich and were used without further purification. They were vaporized in a heated nozzle (held near 370 K) attached to a solenoid-driven pulsed valve (0.8 mm diameter, Parker General Valve, Series 9) with neon and argon, respectively, at 3 bar backing pressure, used as carrier gases creating a supersonic expansion into the vacuum chamber with a rotational temperature of approximately 2 K. The use of neon or argon is just due to logistic circumstances in our lab. For each gas pulse, the ensemble of molecules was polarized with a series of 8 microwave chirps of 4 $\mu$s length varying linearly in frequency from 2 to 8 GHz. The chirped pulse was generated with an arbitrary waveform generator (Tektronix AWG 70002A), amplified to 200 W with a travelling wave tube amplifier and broadcast perpendicular to the propagation of the jet expansion through a horn antenna. A molecular transient emission spanning 40 $\mu$s is then detected through a second horn and amplified by a low noise MW signal amplifier. A total of 4300k FIDs were co-added and Fourier transformed with a Kaiser-Bessel window function to give the broadband rotational spectrum in the frequency domain.  The spectral resolution is better than 10\,kHz and given the signal to noise ratio observed, the frequency measurements have an estimated accuracy of 5\,kHz.

\section{Results and Discussion}

\subsection{Rotational spectra analysis}
\label{results}

 Geometry optimization calculations for both 1-ETN and 2-ETN isomers (see Fig. \ref{moles}) have been reported very recently  before by \citet{Ye2022}. In this work, the authors report accurate molecular structures for both 1-ETN and 2-ETN isomers (see Fig. \ref{moles}) and provide theoretical values for the rotational and centrifugal constants together with values for the electric dipole moment components. The molecular constants for both isomers are shown in Table \ref{const_parent}. We have used these theoretical constants as starting point in the rotational spectra analysis. Both species display no very large $a$ and $b$ dipole moment components, $\mu_a$=0.58~D and $\mu_b$=0.33~D for 1-ETN and $\mu_a$=0.90~D and $\mu_b$=0.11~D for 2-ETN., and null $c$ dipole moment component due to their $C_s$ symmetry.

\begin{figure}
\centering
\includegraphics[angle=0,width=0.5\textwidth]{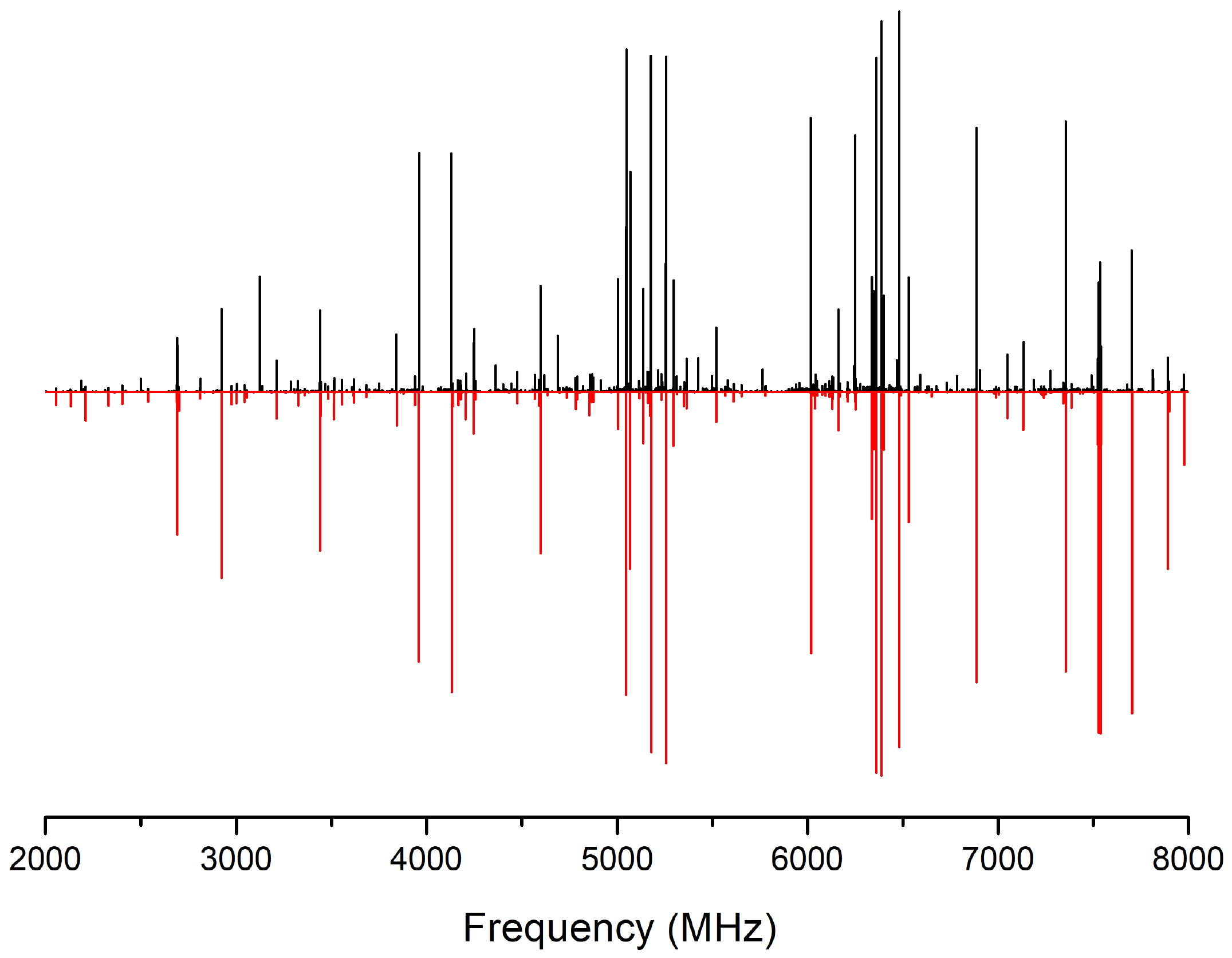}
\caption{Rotational spectrum of 1-ETN in the 2–8 GHz frequency range. The black trace is the experimental spectrum, and the red trace is the simulated spectrum of 1-ETN based on the fitted rotational constants at a rotational temperature of 1~K.} \label{spectrum_1}
\end{figure}

\textbf{1-ETN.} In light of the dipole moment components for 1-ETN it is expected that both $a$- and $b$-type spectra for 1-ETN will be weak but observable. The recorded broadband rotational spectrum in the 2–8 GHz frequency region for 1-ETN is shown in Fig.~\ref{spectrum_1}. On first inspection the broadband rotational spectrum is dense with a strong characteristic $a$-type transitions pattern. In addition, $b$-type lines were also observed, but with lower intensity. Finally, a total of 160 rotational transitions, list included in the Supporting Information, were assigned for 1-ETN. The best fit for all these lines was obtained using the $S$-reduction of Watson's Hamiltonian in the $III^r$ representation \citep{Watson1977}. This fit was done using the Kisiel’s ASFIT program \citep{Kisiel2019} which is best suited when the $S$-reduction in the $III^r$ representation is required. The list of determined constants includes the three rotational constants and the five quartic distortion constants. They are shown in Table~\ref{const_parent} together with the predicted values from \citet{Ye2022}. The calculated rotational constants are in excellent agreement with the experimental values with relative errors ranging from 0.02 to 0.05~\%, indicating that the calculated geometry is very accurate. The discrepancies between experiment and theory are even smaller than those reported for 1-cyanonaphthalene \citep{Ye2022}. The discrepancies between the experimental and predicted values for the quartic distortion constants are larger, as expected, but the agreement can be considered satisfactory as well.

\begin{table*}
\begin{center}
\caption[]{Spectroscopic constants for the two isomers of ethynylnaphthalene.}
{\label{const_parent}
\setlength{\extrarowheight}{1.5pt}
\begin{tabular}{lcccc}
\hline
\hline
&\multicolumn{2}{c}{1-ETN} &\multicolumn{2}{c}{2-ETN}  \\
\cmidrule(lr){2-3} \cmidrule(lr){4-5}
Constants/Units & Exp. & Calc.$^a$ & Exp. & Calc.$^a$ \\
\hline
$A$/MHz                  & 1474.99376(15)$^b$  &  1475.332   &    2713.63397(37)  &    2709.967  \\
$B$/MHz                  &  953.707853(95)  &  953.269	  &    606.040212(41)     &    606.269   \\
$C$/MHz                  &  579.26886(10)   &  579.129	  &    495.478949(41)     &    495.489   \\
$D_J$/kHz                &    0.0593(14)    &  0.0327 	  &    [0.00488]$^c$      &    0.00488  \\
$D_{JK}$/kHz             &   -0.0915(15)    &  -0.0837	  &    [0.00142]          &    0.00142  \\
$D_{K}$/kHz              &    0.0383(18)    &  0.0553 	  &    [0.465]            &    0.465     \\
$d_1$/kHz                &   -0.01309(71)   &  -0.0189 	  &      [-0.00134]       &    -0.00134 \\
$d_2$/kHz                &   -0.01989(30)   &  -0.0130 	  &      [-0.00021]       &    -0.00021 \\
$\sigma_{rms}$/kHz       &          3.2     &             &    3.6             &              \\
$N_{lines}$              &          167     &             &    78              &              \\
$J_{min}$/$J_{max}$      &          0/11    &             &    0/13            &              \\
$K_{a,min}$/$K_{a,max}$  &          0/7     &             &    0/5             &              \\
\hline
\end{tabular}
}
\end{center}
\begin{tablenotes}
      \small
      \item $Notes$. $^a$  \citealt{Ye2022} $^b$ Numbers in parentheses represent the derived uncertainty ($1\,\sigma$) of the parameter in units of the last digit. $^c$ Numbers in brackets have been fixed to the theoretical values.\\
    \end{tablenotes}
\end{table*}

A deeper inspection of the spectrum  allowed us to identify the twelve $^{13}$C isotopic species in natural abundance. The observed number of transitions for each isotopic species vary between 21 and 40, including $a$- and $b$-type transitions. The analysis of the measured transitions was carried out in the same manner than that for the 1-ETN parent species with the exception that the quartic centrifugal distortion constant values were kept fixed to those determined for the parent species. The results obtained from our analysis are shown in Table~\ref{iso_const_1etn} and a list of all the frequency transitions measured are collected in the Supporting Information.

\begin{figure}
\centering
\includegraphics[angle=0,width=0.5\textwidth]{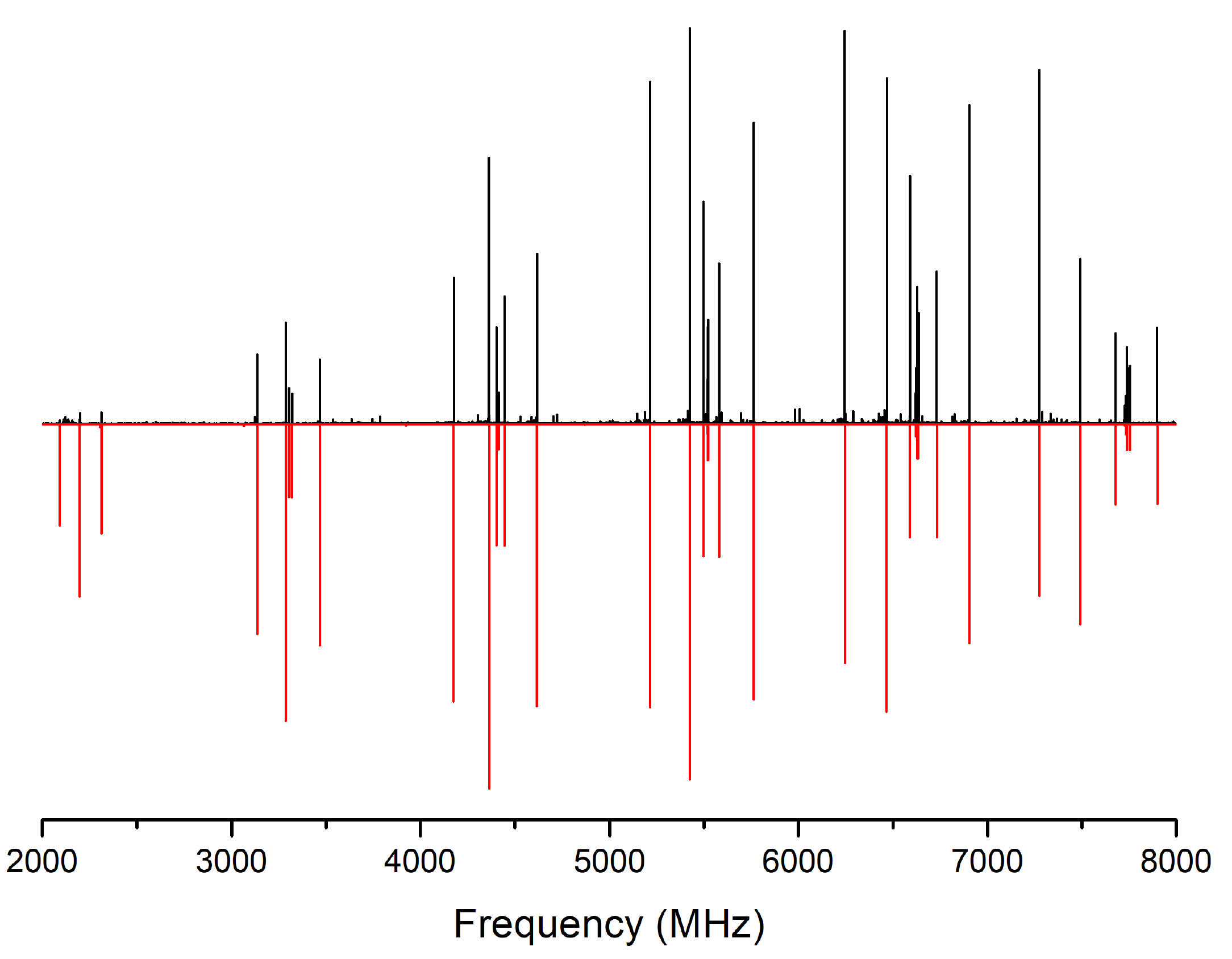}
\caption{Rotational spectrum of 2-ETN in the 2–8 GHz frequency range. The black trace is the experimental spectrum, and the red trace is the simulated spectrum of 2-ETN based on the fitted rotational constants at a rotational temperature of 1~K.} \label{spectrum_2}
\end{figure}

\textbf{2-ETN.}  As it occurs for the 1-ETN isomer, the quantum chemical calculations predict no null $a$- and $b$-dipole moment components for 2-ETN. However, the $\mu_a$ component (0.90~D) is much larger than the $\mu_b$ component, 0.11~D. The observed broadband rotational spectrum for 2-ETN in the 2–8 GHz frequency region, shown in Fig.~\ref{spectrum_2}, is less populated than that for 1-ETN. The main reason is the low value of the $\mu_b$ dipole moment component, which considerably affect to the intensity of the $b$-type spectrum, compared to that for 1-ETN. As mentioned above, the experiments for the 2-ETN were carried out using Ar instead of Ne. However, no significant differences in terms of spectral intensities were encountered between both experiments. In total, we measured a total of 78 rotational transitions for 2-ETN, including 50 $a$-type and 28 $b$-type transitions. The complete list of the assigned transitions is available at the Supporting Information. The best fit for the observed transitions was achieved utilising Watson’s $S$-reduced asymmetric rotor Hamiltonian in the $I^r$ representation \citep{Watson1977} using the SPFIT program \citep{Pickett1991}. In contrast to the 1-ETN species, only the rotational constants could be determined, while the quartic distortion constants were kept fixed to the theoretical values, see Table~\ref{const_parent}. As for the 1-ETN isomer, the experimental rotational constants agree very well with those predicted by \citep{Ye2022}, except the $A$ constant which show larger discrepancies. In this manner, $B$ and $C$ constants show relative errors of 0.04 and 0.002~\%, respectively, while $A$ constant differs from the predicted value by 0.14~\%. 

As for 1-ETN, all the $^{13}$C isotopic species of 2-ETN were observed in natural abundance. Only $a$-type transition could be observed, due to weakness of the $b$-type spectrum. The analysis of the measured transitions was carried out as for the parent species and only the rotational constants were determined. As for the 2-ETN parent isotopologue, the quartic centrifugal distortion constant values were kept fixed to those obtained by quantum chemical calculations. The results obtained from our analysis are shown in Table~\ref{iso_const_2etn} and a list of all the measured frequency is collected in the Supporting Information.

\begin{table*}
\begin{center}
\begin{small}
\caption[]{Spectroscopic constants for the $^{13}$C isotopic species of 1-ETN.}
{\label{iso_const_1etn}
\setlength{\extrarowheight}{1.5pt}
\begin{tabular}{lcccccc}
\hline
\hline
Constants/Units          & $^{13}$C$_1$    & $^{13}$C$_2$   & $^{13}$C$_3$   & $^{13}$C$_4$   & $^{13}$C$_5$   &   $^{13}$C$_6$ \\
\hline
$A$/MHz                  & 1475.00592(39)  & 1466.71741(45) & 1449.91735(47) & 1455.56200(42) & 1472.33576(47) & 1474.09337(46)  \\
$B$/MHz                  &  950.50826(19)  &  948.26618(22) &  952.59009(29) &  953.12640(17) &  951.91083(26) &  943.57085(19)  \\
$C$/MHz                  &  578.08900(10)  &  575.98526(11) &  574.95303(15) &  576.034958(93)&  578.19622(14) &  575.37709(11)  \\
$\sigma_{rms}$/kHz       &      3.5        &       3.9      &        4.9     &      3.2       &      4.3       &     4.6         \\
$N_{lines}$              &      37         &       39       &        39      &      35        &      38        &     42          \\
$J_{min}$/$J_{max}$      &      1/6        &       1/6      &        1/6     &      1/6       &      1/6       &     1/7         \\
$K_{a,min}$/$K_{a,max}$  &      0/3        &       0/4      &        0/4     &      0/3       &      0/4       &     0/3         \\
\hline
\hline
Constants/Units          & $^{13}$C$_7$     & $^{13}$C$_8$    & $^{13}$C$_9$   & $^{13}$C$_{10}$& $^{13}$C$_{11}$& $^{13}$C$_{12}$ \\
\hline
$A$/MHz                  &  1472.10068(54)  &  1460.14552(53) & 1464.07629(65) & 1474.7433(11)  & 1471.19265(46) & 1461.50609(51)  \\
$B$/MHz                  &   939.77460(18)  &   947.52523(25) &  953.27127(27) &  953.73834(55) &  944.00278(19) &  935.73755(19)  \\
$C$/MHz                  &   573.66045(10)  &   574.69570(13) &  577.41815(14) &  579.24303(24) &  575.09433(11) &  570.542557(96) \\
$\sigma_{rms}$/kHz       &     4.1          &        3.9      &     4.1        &     6.8        &       2.9      &     3.3         \\
$N_{lines}$              &     38           &        34       &     37         &     23         &       34       &     35          \\
$J_{min}$/$J_{max}$      &     1/6          &        1/6      &     1/6        &     1/6        &       1/6      &     1/6         \\
$K_{a,min}$/$K_{a,max}$  &     0/3          &        0/3      &     0/3        &     0/3        &       0/3      &     0/3         \\
\hline                                                                                                                     \end{tabular}
}
\end{small}
\end{center}
\begin{tablenotes}
      \small
      \item $Notes$. Numbers in parentheses represent the derived uncertainty ($1\,\sigma$) of the parameter in units of the last digit. In all the fits the quartic distortion constants have been kept fixed to the values determined for the parent species.\\
    \end{tablenotes}
    \end{table*}

\begin{table*}
\begin{center}
\begin{small}
\caption[]{Spectroscopic constants for the $^{13}$C isotopic species of 2-ETN.}
{\label{iso_const_2etn}
\setlength{\extrarowheight}{1.5pt}
\begin{tabular}{lcccccc}
\hline
\hline
Constants/Units         & $^{13}$C$_1$    & $^{13}$C$_2$   & $^{13}$C$_3$   & $^{13}$C$_4$   & $^{13}$C$_5$   &$^{13}$C$_6$    \\
\hline
$A$/MHz                 &   2700.586(41)  &  2713.729(45)  &  2685.692(38)  &  2669.111(37)  &  2704.804(31)  &  2694.9441(30)  \\
$B$/MHz                 &  605.56726(32)  & 603.69734(29)  & 604.53161(30)  & 606.03543(32)  & 605.45232(23)  & 602.288833(22)  \\
$C$/MHz                 &  494.72788(30)  & 493.91390(27)  & 493.53457(28)  & 493.97132(30)  & 494.79014(22)  & 492.348782(22)  \\
$\sigma_{rms}$/kHz      &      4.1        &       3.8      &        3.9     &      4.2       &      4.9       &     4.4        \\
$N_{lines}$             &      22         &       23       &        24      &      22        &      20        &     21         \\
$J_{min}$/$J_{max}$     &      2/7        &       2/7      &        2/7     &      2/7       &      2/7       &     2/7        \\
$K_{a,min}$/$K_{a,max}$ &      0/3        &       0/3      &        0/3     &      0/3       &      0/3       &     0/3        \\
\hline
\hline
Constants/Units         & $^{13}$C$_7$    & $^{13}$C$_8$   & $^{13}$C$_9$   & $^{13}$C$_{10}$& $^{13}$C$_{11}$&$^{13}$C$_{12}$ \\
\hline
$A$/MHz                 &   2713.247(37)  &  2693.173(35)  &  2678.321(42)  &  2708.608(45)  &  2711.924(37)  &  2707.840(35)  \\
$B$/MHz                 &  598.44889(25)  & 599.99403(25)  & 604.24577(33)  & 605.83767(37)  & 598.74852(28)  & 592.54743(27)  \\
$C$/MHz                 &  490.38143(24)  & 490.75654(26)  & 493.09641(31)  & 495.17796(35)  & 490.53742(26)  & 486.23501(25)  \\
$\sigma_{rms}$/kHz      &      3.5        &       3.7      &        4.4     &      4.9       &      3.6       &     3.9        \\
$N_{lines}$             &      21         &       22       &        25      &      25        &      23        &     23         \\
$J_{min}/J_{max}$       &      2/7        &       2/7      &        2/7     &      2/7       &      2/7       &     2/7        \\
$K_{a,min}$/$K_{a,max}$ &      0/3        &       0/3      &        0/3     &      0/3       &      0/3       &     0/3        \\
\hline
\end{tabular}
}
\end{small}
\end{center}
\begin{tablenotes}
      \small
      \item $Notes$. Numbers in parentheses represent the derived uncertainty ($1\,\sigma$) of the parameter in units of the last digit. In all the fits the quartic distortion constants have been kept fixed to the theoretical values reported by \citealt{Ye2022}.\\
    \end{tablenotes}
    \end{table*}

\begin{figure*}
\centering
\includegraphics[angle=0,width=1.0\textwidth]{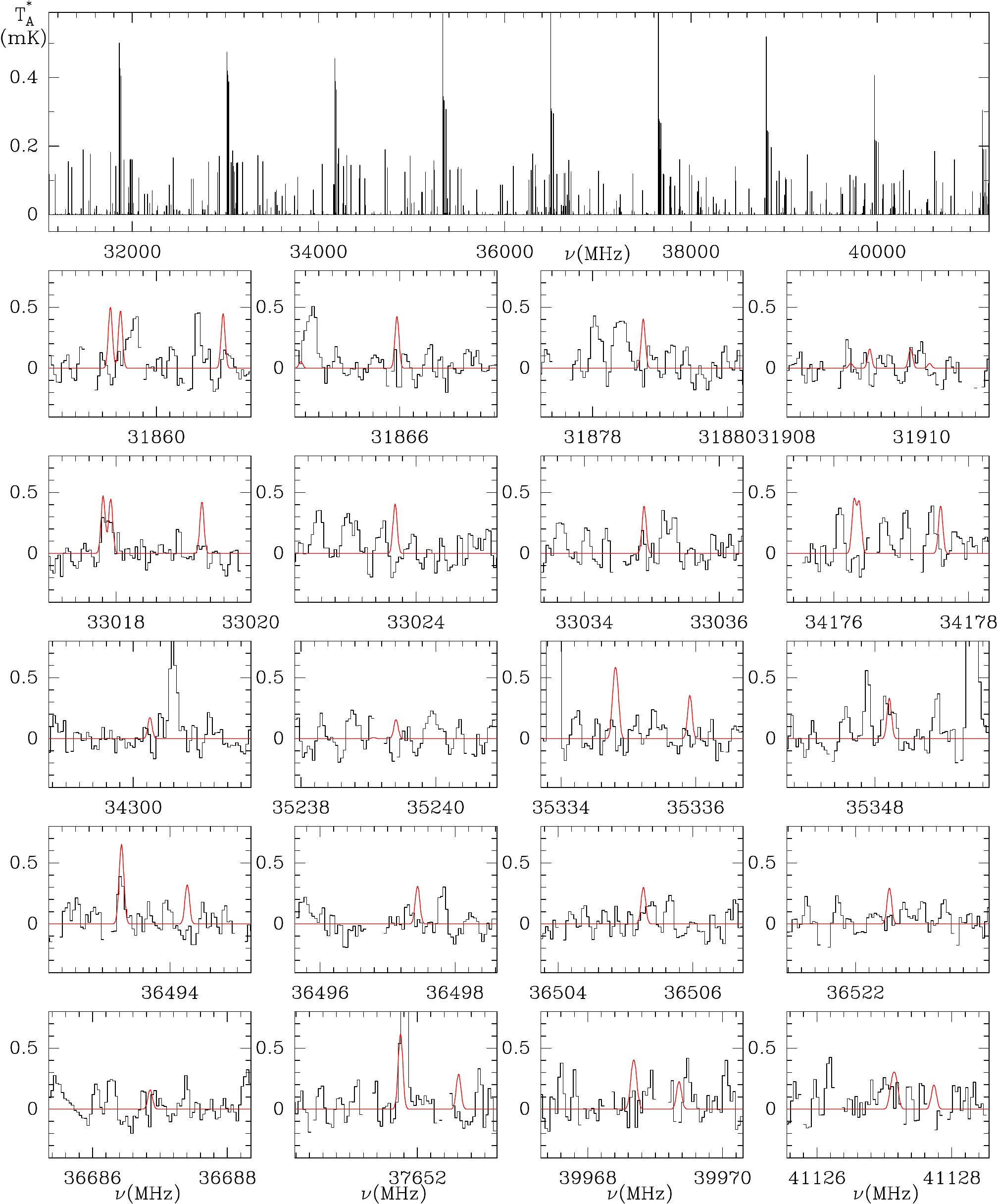}
\caption{The top panel shows the computed rotational spectrum of
1-ETN in the range 31.1-41.2 GHz. We have adopted a rotational temperature
of 10\,K, a line width of 0.6\,km\,s$^{-1}$ and a column density of
1$\times$10$^{13}$ cm$^{-2}$. The other
panels show the QUIJOTE data at the frequency of selected rotational lines of the 1-ETN
isomer. The computed synthetic spectrum is shown in red. Abcisa is frequency and ordinate
is antenna temperature in milli Kelvins.
Blanked channels correspond to negative features produced in the folding of
the frequency switching observations.
} \label{1-etn-tmc1}
\end{figure*}

\begin{figure*}
\centering
\includegraphics[angle=0,width=1.0\textwidth]{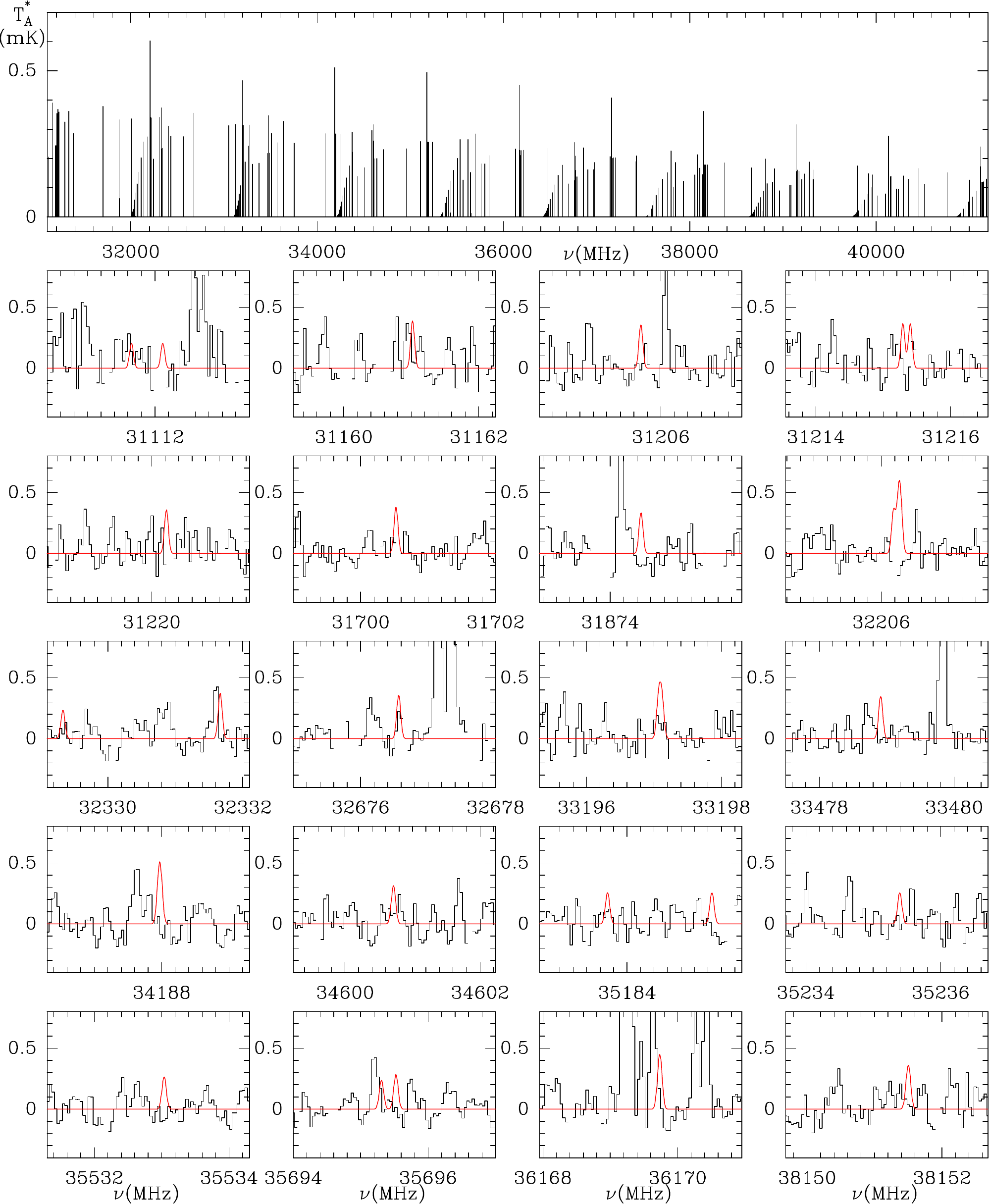}
\caption{Same as Figure \ref{1-etn-tmc1} but for 2-ETN.}
\label{2-etn-tmc1}
\end{figure*}

\subsection{Astronomical search for ethynylnaphthalene}

The observational data used in this article consist of spectra of \mbox{TMC-1} taken with the Yebes 40m telescope towards the cyanopolyyne peak of \mbox{TMC-1}, $\alpha_{J2000}=4^{\rm h} 41^{\rm  m} 41.9^{\rm s}$ and $\delta_{J2000}=+25^\circ 41' 27.0''$. The observations are part of the ongoing QUIJOTE line survey \citep{Cernicharo2021a} carried out during different observing runs between November 2019 and May 2022. QUIJOTE has reached now a level of sensitivity (0.12-0.25 mK across the Q-band) that permits to detect new isotopologues and CN and CCH derivatives of species with abundances of a few $\sim$10$^{-10}$.

The observations were performed using the frequency-switching mode with a frequency throw of 10 MHz in the very first observing runs, in November 2019 and February 2020, 8 MHz in the observations of January-November 2021, and 10 MHz again in the last observing run that took place between October 2021 and May 2022. The total on-source telescope time is 546 h in each polarization (twice this value after averaging the two polarizations), which can be split into 293 and 253 hours for the 8 MHz and 10 MHz frequency throws. The QUIJOTE line survey uses a 7 mm receiver covering the Q band (31.0-50.3 GHz) with horizontal and vertical polarizations. Receiver temperatures in 2019 and 2020 varied from 22 K at 32 GHz to 42 K at 50 GHz. In 2021, some power adaptation carried out in the down-conversion chains changed the receiver temperatures to 16\,K at 32 GHz and 25\,K at 50 GHz. The backends are 16 Fourier transform spectrometers, which provide a bandwidth of 8\,$\times$\,2.5 GHz in each polarization, thus covering practically the whole Q band, with a spectral resolution of 38.15 kHz. The system is described in detail by \citet{Tercero2021}.

The intensity scale used is antenna temperature, $T_A^*$, which is calibrated using two absorbers at different temperatures and the ATM package \citep{Cernicharo1985,Pardo2001}. Calibration uncertainties were assumed to be 10\% based on the observed repeatability of
the line intensities between different observing runs. All data were analysed using the software  GILDAS\footnote{\texttt{http://www.iram.fr/IRAMFR/GILDAS/}}.

\begin{table}
\begin{center}
\caption[]{Rotational partition function for 1-ETN and 2-ETN at different temperatures.}
\scalebox{1}{
\label{pfunction}
\begin{tabular}{ccc}
\hline
\hline
Temperature/K   &\hfill 1-ETN\hfill\mbox{} &\hfill 2-ETN\hfill\mbox{}\\
\hline
300.000  &  971140.1  &   970842.2  \\
225.000  &  630707.4  &   630564.7  \\
150.000  &  343252.0  &   343199.5  \\
 75.000  &  121340.5  &   121330.2  \\
 37.500  &   42901.2  &    42898.4  \\
 18.750  &   15170.8  &    15169.6  \\
  9.375  &    5366.2  &     5365.4  \\
\hline
\end{tabular}
}
\end{center}
\end{table}

The frequency predictions for both isomers 1-ETN and 2-ETN were implemented in the MADEX code \citep{Cernicharo2012}, to compute the synthetic spectrum of both isomers in the Q-band we have assumed a rotational temperature of 10\,K, a line width of 0.6 km\,s$^{-1}$, and a v$_{LSR}$ of 5.83 km\,s$^{-1}$ \citep{Cernicharo2020a}. We used the dipole moment components reported by \citep{Ye2022} and shown in Section~\ref{results}. The partition functions employed in the frequency predictions are shown in Table \ref{pfunction}. They were calculated using the SPCAT program \citep{Pickett1991} and at maximum value of $J$=300. We focused our search on the $a$- and $b$-type $R$-branch transitions with $K_a$ values 0, 1 and 2 which are degenerated in frequency increasing the intensity of the expected lines. None of these series of lines are clearly seen in our TMC-1 data. The sensitivity of the QUIJOTE line survey varies between 0.12 and 0.25\,mK in the 31-50\,GHz domain. Figures \ref{1-etn-tmc1} and \ref{2-etn-tmc1} show the predicted spectrum of both isomers across the Q-band and the comparison of the strongest predicted lines with the data. Frequency predictions for 1-ETN have uncertainties between 10-60 kHz in the Q-band, while those of 2-ETN have larger uncertainties due to the impossibility to get the distortion constants from the laboratory data. Although in some cases the predicted lines agree with features in the QUIJOTE data, most lines are missing. We tried to assign some of the spectral features appearing close to the predicted lines, but we fail in getting a reasonable fit when merging them with the laboratory data. Adopting the observed 3$\sigma$ limits to the intensity of strongest lines, we derive a 3$\sigma$ upper limit to the column density of 1-ETN in TMC-1 of 1.0$\times$10$^{13}$ cm$^{-2}$. For the 2-ETN isomer, which have a larger $\mu_a$ dipole moment the derived upper limit is smaller, 6.8$\times$10$^{12}$ cm$^{-2}$.

It is interesting to compare the abundance of the ethynyl and cyano derivatives of naphtalene. \citet{McGuire2021} derived column densities for 1-cyano and 2-cyanonaphtalene following a stacking spectral method and fitting four velocity components with different sizes. The total column density of 1-cyano and 2-cyano naphtalene is 7.35$\times$10$^{11}$ cm$^{-2}$ and 7.04$\times$10$^{11}$ cm$^{-2}$ respectively. Hence, the 3$\sigma$ upper limits to the abundance ratio of ethynyl over cyano derivatives of naphtalene are 13.6 and 9.7 for the positions 1 and 2 of the derivatives. The significant lower dipole moments of the
ethynyl derivatives compared to the cyano ones could explain that only upper limits are obtained. By stacking the non-blended lines of 1- and 2-ETN, these upper limits could be improved by a factor of 2 giving values for the abundance ratios similar to those found for the ethynyl and cyano derivates of cyclopentadiene, $\sim$7 \citep{Cernicharo2021c}. We note, however, that for smaller hydrocarbons such as C$_2$H$_4$ the CH$_2$CHCCH/CH$_2$CHCN abundance ratio is $\sim$2 \citep{Cernicharo2021d}. Future versions of the QUIJOTE line survey, together with additional laboratory data at higher frequencies for the ethynyl derivatives, could certainly improve the present results and, perhaps, allow the detection of both isomers.

\section{Conclusions}

In this work, we have investigated the pure rotational spectra of the two isomers of the ethynyl derivative of naphthalene. The measurements have been carried out in the 2-8 GHz frequency region using a high-resolution broadband spectrometer. From the analysis of the rotational spectra we obtained accurate rotational parameters for both isomers which allowed us to obtain frequency predictions in the Q-band frequency range and search for them using the QUIJOTE line survey. Although they were not detected towards TMC-1 we obtained upper limits for both species.

\section*{Acknowledgements}

C. C. and J. C. thank  European Research Council for funding support under Synergy Grant ERC-2013-SyG, G.A. 610256 (NANOCOSMOS) and Ministerio de Ciencia e Innovación through the PID2019–107115GB–C21 project. I.P acknowledges Ministerio de Ciencia e Innovación (Grant PID2020–117925GA–I00) and Junta de Castilla y León (Grant INFRARED-FEDER IR2020-1-UVa02). We thank the referee for a careful reading of the manuscript and for useful comments.

C.C. conceived the study, performed the analysis of the rotational spectroscopy experiments, and wrote the original draft. I.P is responsible for the resources and funding and carried out all the laboratory experiments. J.C. performed the analysis of all TMC-1 surveys and carried out the astronomical search of the observed molecules. He is also responsible for the resources and funding for this investigation. All the authors reviewed and commented the manuscript.

\section*{Data Availability}

The data underlying this article will be shared on reasonable request to the corresponding authors.

\bsp	
\label{lastpage}
\end{document}